\documentclass[twocolumn]{aastex631}

\usepackage[T1]{fontenc}
\usepackage{graphicx}	% Including figure files
\usepackage{amsmath}	% Advanced maths commands
\usepackage{amssymb}	% Extra maths symbols
\usepackage{acronym}

\acrodef{PE}{Parameter Estimation}
\acrodef{GW}{gravitational wave}
\acrodef{BBH}{binary black holes}
\acrodef{GWTC}{Gravitational Wave Transient Catalog}
\acrodef{BNS}{Binary Neutron Star}
\acrodef{NSBH}{Neutron Star--Black Hole}
\acrodef{LVK}{LIGO Scientific, Virgo and KAGRA Collaborations}

\newcommand{\D}{\mathrm{d}}
\newcommand{\M}{\mathrm{M}}

\newcommand{\checkme}[1]{{#1}}
\newcommand{\peakonemch}{\checkme{8.1}}
\newcommand{\peaktwomch}{\checkme{13.9}}
\newcommand{\peakthreemch}{\checkme{26.9}}
\newcommand{\peakfourmch}{\checkme{45.3}}

\newcommand{\ratem}{\checkme{18.4}}
\newcommand{\rateu}{\checkme{12.6}}
\newcommand{\ratel}{\checkme{8.3}}

\newcommand{\lmratio}{\checkme{0.47}}
\newcommand{\lspinz}{\checkme{-0.41}}
\newcommand{\rspinz}{\checkme{+0.39}}

\newcommand{\lszLomass}{\checkme{-0.41}}
\newcommand{\rszLomass}{\checkme{+0.38}}
\newcommand{\lszHimass}{\checkme{-0.55}}
\newcommand{\rszHimass}{\checkme{+0.62}}

\graphicspath{{./}{figures/}}
\begin{document}

\title{Exploring Features in the Binary Black Hole Population}

\correspondingauthor{Vaibhav Tiwari}
\email{tiwariv@cardiff.ac.uk}

\author{Vaibhav Tiwari}
\affiliation{Gravity Exploration Institute \\
School of Physics and Astronomy, \\ 
Cardiff University, Queens Buildings, The Parade \\ 
Cardiff CF24 3AA, UK.}



\begin{abstract}
Vamana is a mixture model framework that infers the astrophysical distribution of chirp mass, mass ratio, and spin component aligned with the orbital angular momentum for the \ac{BBH} population. We extend the mixing components in this framework to also model the redshift evolution of merger rate and report all the major one and two-dimensional features in the \ac{BBH} population using the 69 \ac{GW} signals detected with a false alarm rate $<1\mathrm{yr}^{-1}$ in the third \ac{GWTC}-3. Endorsing our previous report and corroborating recent report from LIGO Scientific, Virgo, and KAGRA Collaborations, we observe the chirp mass distribution has multiple peaks and a lack of mergers with chirp masses $10 \textrm{--} 12M_\odot$. In addition, we observe aligned spins show mass dependence with heavier binaries exhibiting larger spins, mass ratio shows a dependence on the chirp mass but not on the aligned spin, and the redshift evolution of the merger rate for the peaks in the mass distribution is disparate. These features possibly reflect the astrophysics associated with the \ac{BBH} formation channels. However, additional observations are needed to improve our limited confidence in them. 

\end{abstract}


\renewcommand{\arraystretch}{1.1}

\section{Introduction}

%All papers should start with an Introduction section, which sets the work
%in context, cites relevant earlier studies in the field by \citet{Fournier1901},
%and describes the problem the authors aim to solve \citep[e.g.][]{vanDijk1902}.
%Multiple citations can be joined in a simple way like \citet{deLaguarde1903, delaGuarde1904}.
\ac{LVK} recently released \ac{GWTC}-2.1, a deep extended catalog of observations made during the first half of the third observation run (O3a)~\citep{2021arXiv210801045T}. This was followed by the release of \ac{GWTC}-3, a catalog inclusive of observations made during the second half of the third observation run (O3b)~\citep{o3b_cat}. Observation of \ac{GW} signals have also been reported in previous catalogs \ac{GWTC}-1 and \ac{GWTC}-2~\citep{2019PhRvX...9c1040A, 2021PhRvX..11b1053A}. Advanced LIGO and advanced Virgo~\citep{2015CQGra..32g4001L, 2015CQGra..32b4001A} have now detected a total of 69 \ac{BBH} observations at a false alarm rate of less than once per year.

These observations have begun to probe the \ac{BBH} population and have presented many unexpected surprises. The observation of massive binary, GW150914~\citep{2016PhRvL.116f1102A}, was in contrast to the early on expectation of observing lighter binaries~\citep{1998ApJ...499..367B, 2001ApJ...554..548F, 2010ApJ...725.1918O, 2011ApJ...741..103F}. Observations of multiple low-spin binaries as measured from the \ac{GW}~\citep{2017Natur.548..426F, 2018ApJ...854L...9F, 2018ApJ...868..140T} contrasts the high spin black hole companion in various x-ray binary measurements~\citep{gou2011, Miller:2009cw, mcclintock-2006-652, McClintock:2011zq, mm2015, 2021arXiv211102935F}. And, an interesting and somewhat surprising feature in the \ac{BBH} population is an emerging structure in the mass distribution. The observations tend to cluster around multiple peaks. In addition, there is a gap in the chirp mass distribution lacking mergers in the range 10--12$M_\odot$~\citep{2021ApJ...913L..19T, o3b_rnp}.

Observations have further expanded on these features, indicating that lighter binaries contribute significantly to the total merger rate~\citep{2019PhRvX...9c1040A, 2021PhRvX..11b1053A, 2021arXiv210801045T, o3b_cat}, black holes in heavier binaries tend to have larger spin magnitude~\citep{2021ApJ...913L..19T,2021arXiv210902424G, Hoy:2021rfv}, and as reported in this article the early results indicate the merger rate evolves with redshift, but, at different rate for the peaks in the mass distribution.

Periodic increase in the number of observations has motivated multiple reports on the \ac{BBH} population~\citep{2018ApJ...856..173T, 2019PhRvD.100d3012W, 2019ApJ...882L..24A, 2019MNRAS.484.4216R, 2021ApJ...913L...7A, 2021PhRvD.104h3010R,o3b_rnp}. In this article, we report on the features in the \ac{BBH} population inferred by the mixture model framework Vamana~\citep{2021CQGra..38o5007T} using the \ac{GW} observations detected in \ac{GWTC}-3. This article is laid out as follows: We briefly discuss the analysis in Section~\ref{sec:analysis}, the features in the predicted population in Section~\ref{sec:popprop}, some features in the context of hierarchical merger scenario in Section~\ref{sec:hierarchy} and astrophysical implication in Section~\ref{sec:hiermerge}.

\section{Data Selection and Analysis}
\label{sec:analysis}
Gravitational wave observations made in the last three observation runs have been reported over multiple catalogs~\citep{2019PhRvX...9c1040A, 2021PhRvX..11b1053A, 2021arXiv210801045T, o3b_cat}. We analyse the \ac{BBH} mergers reported with a false alarm rate of at most once per year. Independent searches have reported additional \ac{GW} observations~\citep{2019ApJ...872..195N, 2020ApJ...891..123N, 2019PhRvD.100b3007Z, 2020PhRvD.101h3030V} but we leave these observations out due to lack of a framework that can self consistently combine results from independent search analysis. As we restrict our analysis to \ac{BBH}, we exclude binaries that have at least one component consistent with a neutron star. They are, GW170817, GW190425, GW200105, GW190917, GW200105, GW200115~\citep{2017PhRvL.119p1101A, 2020ApJ...892L...3A, 2021ApJ...915L...5A, o3b_cat}. Finally, we also exclude GW190814~\citep{2020ApJ...896L..44A}, which has secondary mass substantially different from the remaining \ac{BBH} observations ($\sim 2.6 M_{\odot}$) and its exclusion is not expected to impact the inference on the bulk \ac{BBH} population~\citep{2021arXiv210900418E}. The total number of selected observations is 69. 

We use the mixture model framework Vamana to predict the population~\citep{2021CQGra..38o5007T}. Vamana uses a mixture of components, each composed of a Gaussian, another Gaussian, and a power-law to model the chirp mass, both the aligned spin components, and mass ratio respectively. Similar to multiple previous works, the redshift evolution of merger rate is modeled using a \emph{Single} power-law with exponent $\kappa$ quantifying the merger rate evolution for the full population~\citep{2018ApJ...863L..41F, 2019ApJ...882L..24A, 2020ApJ...896L..32C, 2020PhRvD.102l3022R, 2021ApJ...913L...7A, o3b_rnp},
\begin{equation}
    \mathcal{R}(z) = \mathcal{R}\,(1 + z) ^ \kappa,
    \label{eq:pz}
\end{equation}
where $\mathcal{R}$ is the population averaged merger rate at $z = 0$.
Some features of the population inferred using Vamana for the GWTC-3 observations have already been reported \citep{o3b_rnp}. In this article we report features not presented earlier. In addition, we report inference made using the \emph{Mixed} model created by assigning mixture component independent $\kappa$ values. This extension facilitated separable modeling of the redshift evolution for the merger rate in different regions of parameter space and resulted in the identification of an additional feature at a moderate credibility: the redshift evolution of the merger rate shallows for the fourth peak in the chirp mass distribution.

We use a Jefferey's prior for the \emph{Single} model, defined as \citep{Grigaityte664243},
\begin{equation}
    p(\kappa) \propto \frac{1}{|1 + k|},\quad \mathrm{for}\;|1 + \kappa| \in [0.1, 10],
    \label{eq:jefferey_kappa}
\end{equation}
and assign a uniform prior for $\kappa$ between -1.1 and -0.9 such that distribution is piecewise continuous. For the \emph{Mixed} analysis we facilitate separable modeling by letting components choose uniformly between $\kappa_i - 3$ and $\kappa_i + 3$, for a $\kappa_i$ sampled from the prior distribution in Equation~\ref{eq:jefferey_kappa}. The choice of range is arbitrarily chosen to allow the components to have few orders of magnitude variation in merger rate at a redshift of one. In practice, a bigger interval allows for more variation, but for large values of $\kappa$, inference on population hyper-parameters are inaccurate due to the errors incurred in the importance sampling employed in the Bayesian analysis~\citep{2021CQGra..38o5007T}. Although we use a sufficiently open model, as $\kappa$ values outside this range are also supported by the data, the median and credible intervals presented in this article is subjected to our chosen prior. Our choice on the remaining hyper-parameter priors and their ranges for the \emph{Mixed} model remain unchanged compared to the \emph{Single} model~(please refer to appendix B.1.d in \citet{o3b_rnp} for the complete description).  
Our inference is robust for a wide range of component numbers in the mixture. We use eleven components in the presented analysis as this choice maximises the marginal likelihood. 

The observed binary population is biased compared to the true astrophysical distribution due to the selective sensitivity of the gravitational wave network towards \ac{BBH} masses and spins.  This bias is corrected by estimating the sensitivity of the searches towards simulated signals added to the data set~\citep{2018CQGra..35n5009T}. The large-scale simulation runs performed to estimate this sensitivity has been obtained  using the waveform model \texttt{SEOBNRv4PHM}~\citep{Bohe:2016gbl,2020PhRvD.102d4055O}. To reduce any systematic difference between simulations for sensitivity estimation and \ac{PE} samples of the \ac{GW} signals, we preferentially choose the \ac{PE} samples obtained using the same waveform model, wherever available. However, using combined \ac{PE} samples from \texttt{SEOBNRv4PHM} and \texttt{IMRPhenomXPHM}~\citep{Pratten:2020ceb}, as used in~\citet{o3b_rnp}, has only a small effect on the results. The binary parameters are estimated in the detector frame, to change to the source frame quantities we use the Planck15 cosmology \citep{2016A&A...594A..13P}. All the \ac{PE} samples and simulation campaign's data is publicly available~\citep{losc, ligo-O1O2O3-search-sensitivity}.

Although \emph{Mixed} model introduces ten additional hyper-parameters, the marginal likelihood for both \emph{Single} and \emph{Mixed} models is approximately the same. Unless otherwise noted, all the numerical results reported in this article are for the \emph{Mixed} model with the median as the central value along with the 90\% credible interval.

\section{The Binary Black Hole Population}
\label{sec:popprop}
In this section we discuss the various features in the binary black hole population.
\subsection{Mass Distribution}
\label{sec:mass_distr}
\begin{figure*}
    \centering
    \includegraphics[width=0.97\textwidth]{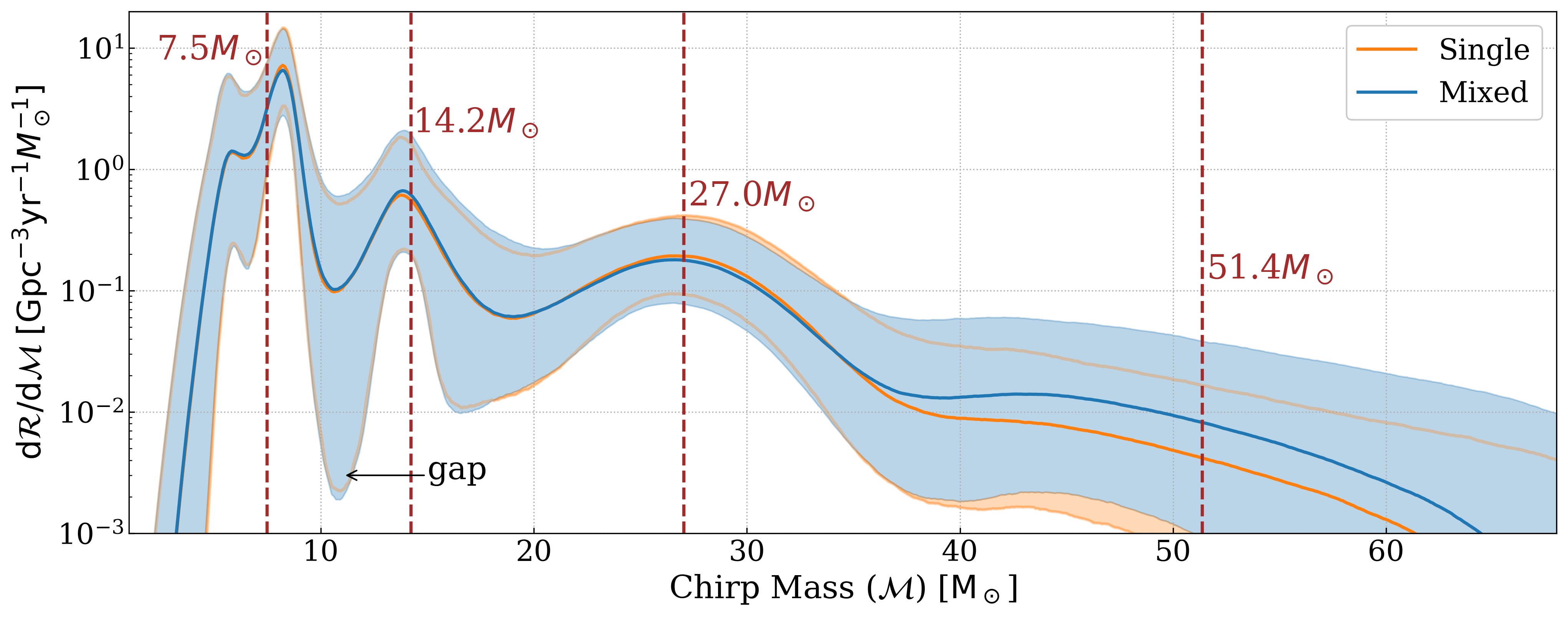}
    
    \vspace{2mm}
    
    \includegraphics[width=0.97\textwidth]{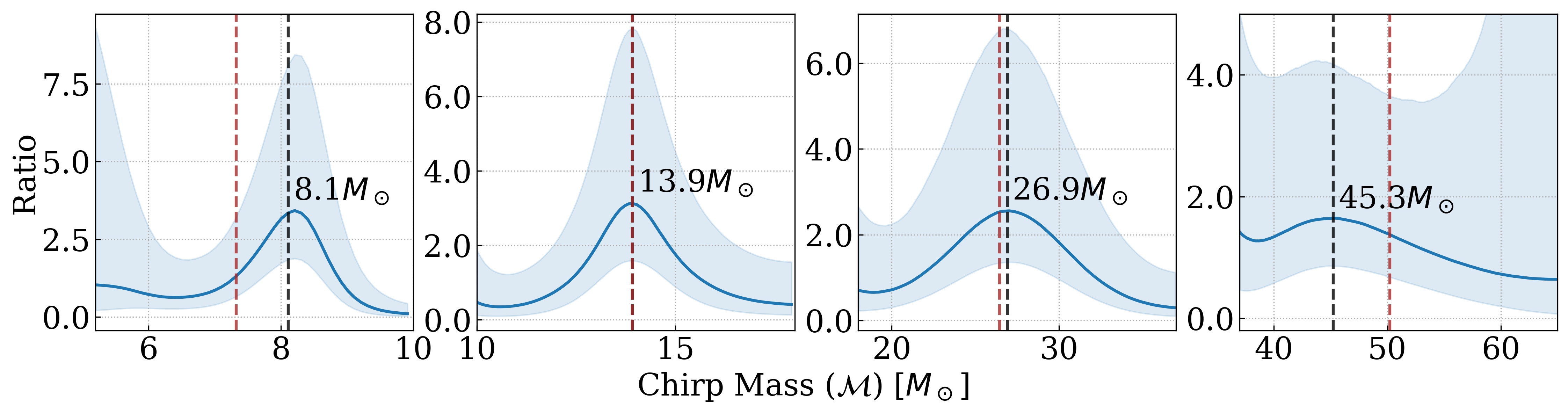}
    \caption{\emph{Top}) The predicted one-dimensional chirp mass distribution at $z=0$. The solid line shows the posterior median and the shaded regions (light solid line when hidden) show the 90\% credible interval. The chirp mass distribution shows presence of a number of peaks. The lack of mergers in the chirp mass range $10-12 M_{\odot}$ is labeled as 'gap'. Starting at an arbitrarily chosen chirp mass value, the dashed brown lines are placed at a factor of 1.9. \emph{Bottom}) The brown lines match well with the location of the peaks. The location of the peaks is the local maxima in the blue curve and indicated by the black dashed line. The blue curve is the median of ratio between the posterior and prior chirp mass density, each truncated and normalised for the shown interval. The shaded region is the 50\% credible region.
    }
    \label{fig:post_mc}
\end{figure*}
\begin{table}[b]
\vspace{1mm}
\begin{ruledtabular}
\begin{tabular}{ccll}
   Peak & Chirp Mass Range & Local Maxima & Credibility\\
   \hline
  1  & 5.2--10 & \peakonemch & 93\\
  2 & 10--18 & \peaktwomch & 88\\ 
  3 & 18--37 & \peakthreemch & 86\\ 
  4 & 37--67  & \peakfourmch & 70\\
\end{tabular}
\end{ruledtabular}
\caption{We define four chirp mass intervals that enclose the peaks. We make random draws from the posterior and prior chirp mass density. For each interval, we truncate and normalise these densities, and calculate the ratio between them. We define a peak at the chirp mass value where this ratio is greater than one at the highest credibility. Here, we define credibility as the percentage of draws with a ratio greater than one at a given mass value. In this table, we summarise the choice on intervals, location of the peaks and our confidence in them. All masses are in $M_\odot$.} 
\label{table:peaks_summary}
\end{table}
%
% In an earlier publication we reported an emerging structure in mass distribution. Those results were based on 39 observations we chose to model in GWTC-2 when our results were based on 39 observations \citep{2021ApJ...913L..19T}. The addition of around 30 observations reported in GWTC-2.1 and GWTC-3 has kept this structure intact \citep{o3b_rnp}. We also report multiple features in the predicted mass, spin, and redshift distribution of the \ac{BBH} population and discuss some of these features in the context of the hierarchical merger scenario, which we suggested could be a possible source of the structure in the mass distribution.
In a previous article, we reported an emerging structure in the mass distribution~\citep{2021ApJ...913L..19T}. Independent analyses have reported similar features since then~\citep{2021-Edelman-PowerlawSpline, 2021arXiv211212659S, 2021ApJ...917...33L, 2021arXiv210513983V, 2021arXiv210905960R}. Our previous population predictions were based on 39 observations reported in \ac{GWTC}-2. The addition of newly reported or previously ignored observations (we used stricter selection criteria in the previous analysis), 30 in number, has kept the structure intact~\citep{o3b_rnp}. The observations cluster around four peaks and there is a lack of mergers in the chirp mass range 10--12 $M_\odot$. The upper panel in Fig.~\ref{fig:post_mc} shows the predicted chirp mass distribution and the lower panel shows the location of the peaks. Each peak occurs at approximately double the mass of the previous peak. We have a substantial confidence in the presence of the first peak, the second and the third peak, and a marginal confidence in the presence of the fourth peak. We quantify the location of the peaks and our confidence in them in Table~\ref{table:peaks_summary}.
\begin{figure*}
    \centering
    \includegraphics[width=0.95\textwidth]{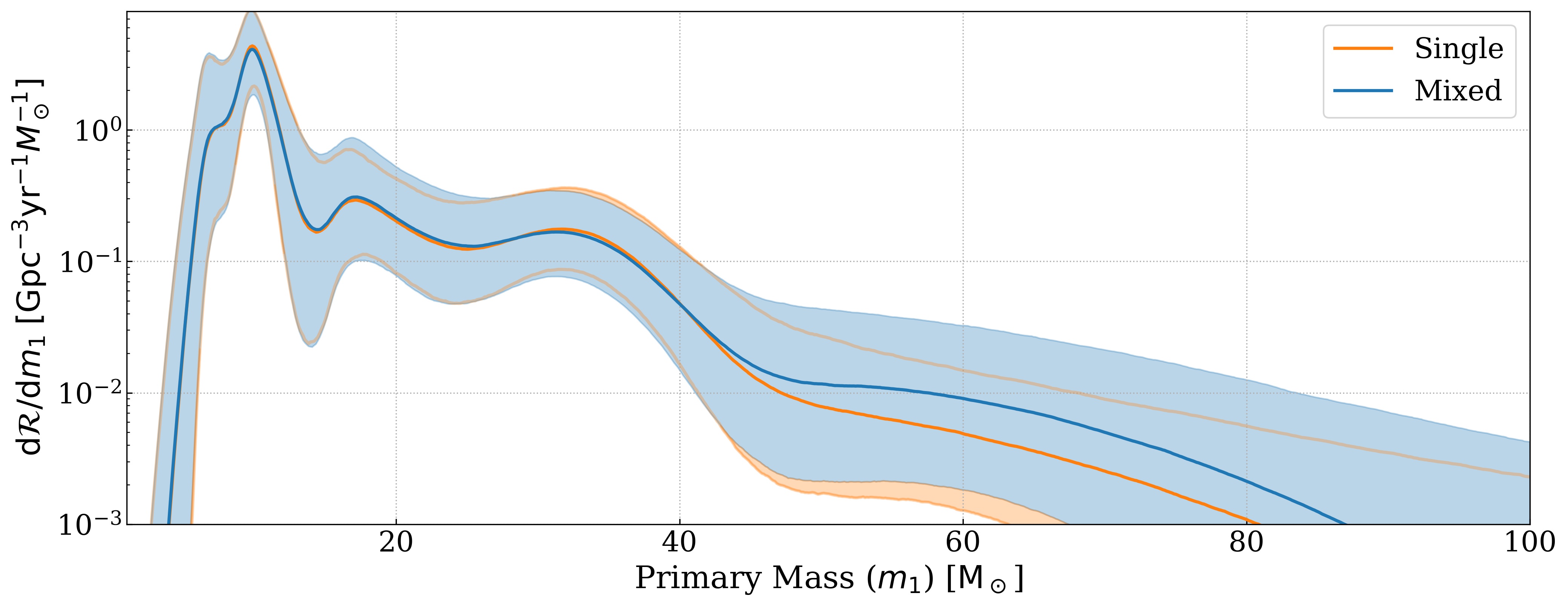}
    \caption{The predicted primary mass distribution for the two redshift models at redshift $z = 0$. The solid line is the median distribution and the shaded region (light solid line when hidden) shows the 90\% credible interval. The primary mass distribution shows a similar structure, and although the locations of the peaks are different compared to their locations on the chirp mass distribution, their relative location bears similar factors. This is understandable as mass ratio distribution shows only a weak dependence on the chirp mass (please see section \ref{sec:correlated}).}
    \label{fig:post_m1}
\end{figure*}

 Fig.~\ref{fig:post_m1} shows the predicted primary mass distribution. A similar structure can be observed. Compared to Fig.~\ref{fig:post_mc}, the peaks are located at different mass values, but the locations bear a similar factor. The features are less pronounced compared to the chirp mass distribution. Possibly, because the mass ratio is not measured precisely. It is also possible that the astrophysical primary mass distribution does not exhibit peak as prominently as the chirp mass distribution, however, that will require a unique combination of primary mass and mass ratio distributions. Although we do not show the component or the secondary mass distributions they exhibit similar structure. We note that the mass of black holes in the three reported neutron star--black hole binaries (GW190917, GW200105, and GW200115) are consistent with the first peak in the primary mass distribution~\citep{2021ApJ...915L...5A, o3b_cat}. We also note that most of the observations, reported in multiple independent searches, but not included in the presented analysis, also follow this clustering~\citep{2020PhRvD.101h3030V, 2019PhRvD.100b3007Z, 2019ApJ...872..195N, 2020ApJ...891..123N, 2021arXiv210509151N}.

\subsection{Spin and Mass Ratio Distribution}
\begin{figure*}
    \centering
    \includegraphics[width=0.95\textwidth]{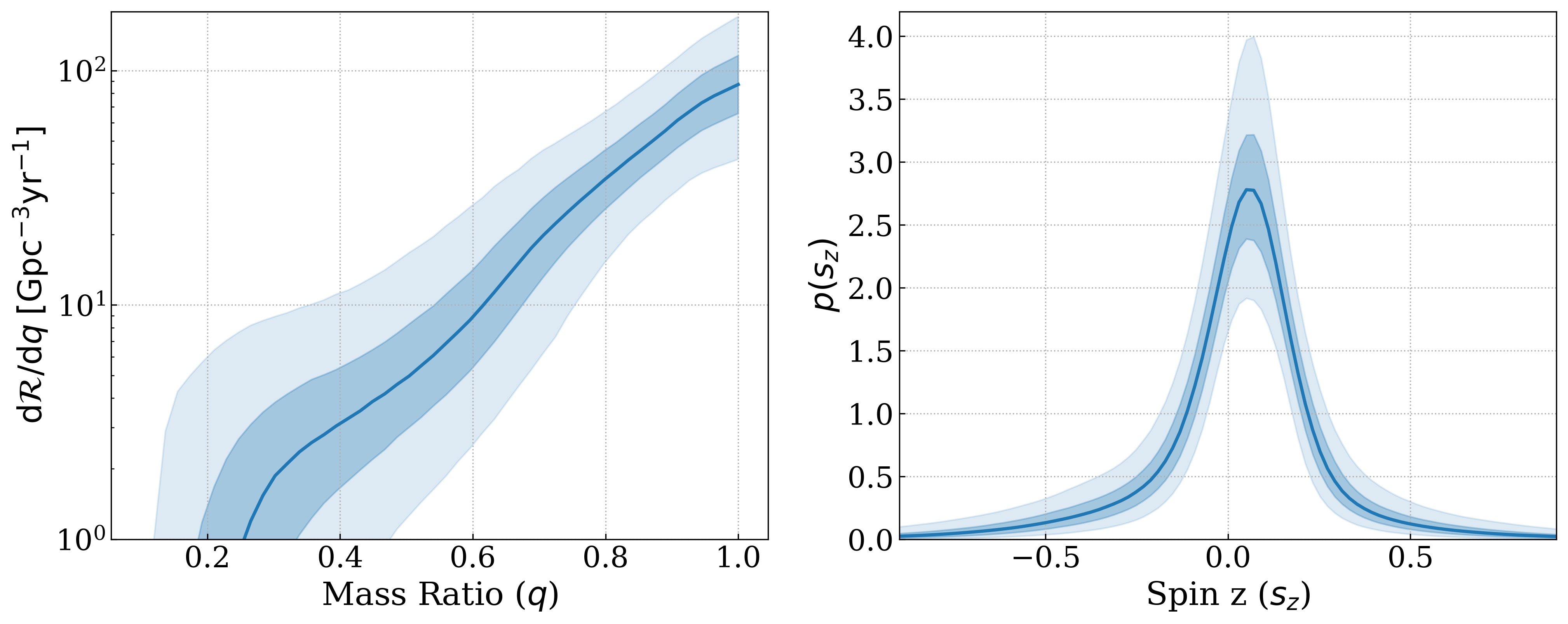}
    \caption{The predicted one-dimensional mass ratio and aligned spin distributions.  \emph{Left}) The solid line is the distribution median and the shaded regions show the 50\% and the 90\% credible interval. The mass ratio distribution is peaked towards equal masses, with 95\% of support above \lmratio. \emph{Right}) The predicted aligned spin distribution shows support for small aligned spins,  with the distribution,  peaked near zero, and  90\%  of the distribution is contained within the range  [\lspinz, \rspinz]. The one-dimensional distribution is dominated by low mass binaries; all of them have been measured with low spin magnitude. The spins show a correlation with the chirp mass which we discuss in section \ref{sec:correlated}.}
    \label{fig:post_qsz}
\end{figure*}

Fig.~\ref{fig:post_qsz} shows the one-dimensional mass ratio and aligned spin distributions obtained by marginalising over other parameters. Similar to results post \ac{GWTC}-2, the mass ratio is well modeled by a decaying power law. Ninety five percent of the binaries have a mass-ratio of greater than half. For the second half of the third observation run there has been an increase in the fraction of binaries that exhibit higher spin magnitudes, thus the aligned distribution has slightly broadened since our last report~\citep{2021ApJ...913L..19T}. 

The choice of the power-law function in modeling the mass ratio will inadvertently impact the measurement of the spins as the two parameters are significantly correlated~\citep{2013PhRvD..87b4035B, 2018ApJ...868..140T}. Due to this correlation, the mass ratio is measured less accurately, thus also impacting the measurement of the component masses. Among the multiple phenomenological distributions we have tested to model the mass ratio, the marginal likelihood is maximised when using a power law.

\subsection{Redshift Evolution of Merger Rate}
\begin{figure*}
    \centering
    \includegraphics[width=0.97\textwidth]{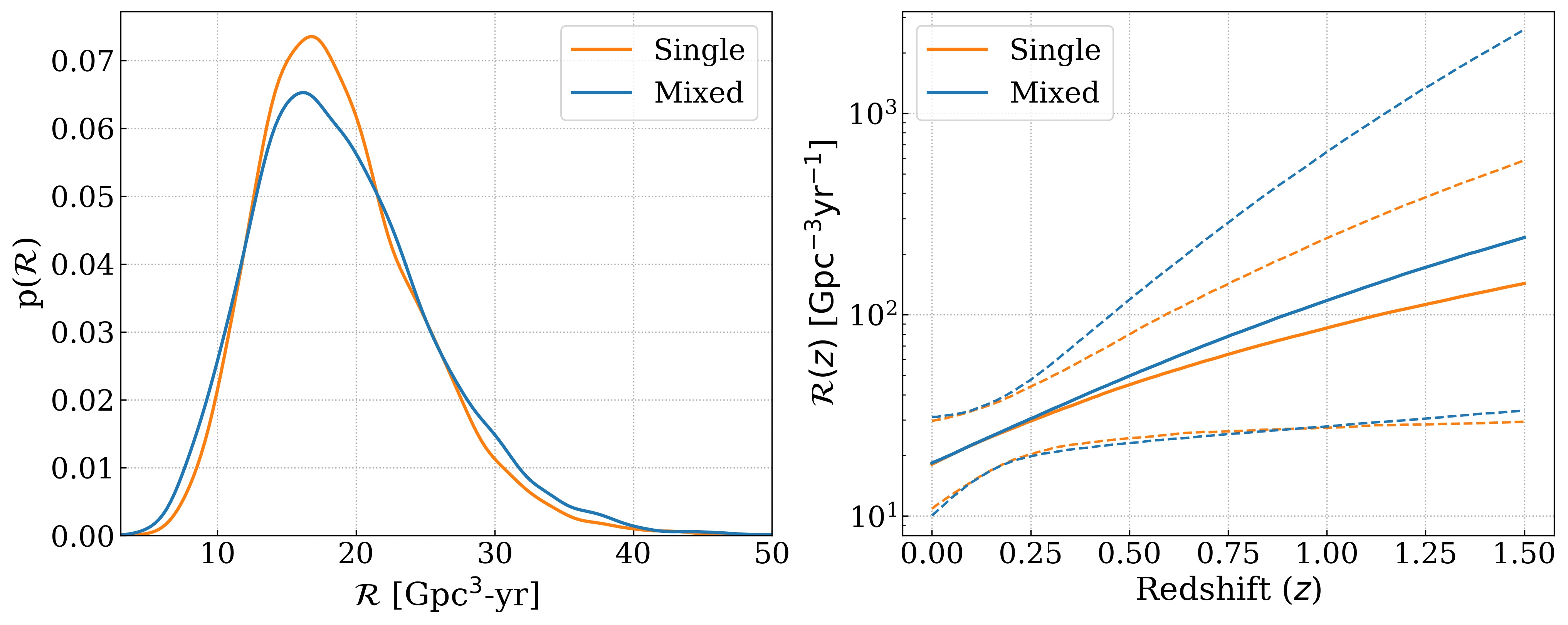}
    \caption{\emph{Left}) The merger rate of binary black-holes in the local universe for the two redshift models. The \ac{BBH} merger rate for the \emph{Mixed} model is $\ratem^{+\rateu}_{-\ratel}\,\mathrm{Gpc}^{-3}\,\mathrm{yr}^{-1}$. \emph{Right}) The redshift evolution of the merger rate. Solid curves are the median distribution and dashed curves enclose the 90\% credible intervals. The \emph{Mixed} model shows a steeper increase in the merger rate.}
    \label{fig:rates}
\end{figure*}

We estimate the population averaged \ac{BBH} merger rate to be $\ratem^{+\rateu}_{-\ratel}\,\mathrm{Gpc}^{-3}\,\mathrm{yr}^{-1}$ at $z=0$. This is consistent with the merger rate reported in~\citet{o3b_rnp}. The merger rate corresponding to chirp mass intervals listed in Table~\ref{table:peaks_summary} are reported in Table~\ref{table:peaks_rate_summary}. The first chirp mass interval contributes around 70\% of the mergers. Fig.~\ref{fig:rates} shows the posterior on the merger rate and its evolution with the redshift. For both the models, the merger rate is increasing with redshift at credibility greater than 95\%. 
\begin{table}[b]
\begin{ruledtabular}
\begin{tabular}{ccccc}
   Interval & 5.2--10 & 10--18 & 18--37 & 37--67\\
   \hline
   $\mathcal{R}$ & $12.8^{+9.5}_{-6.0}$  & $2.4^{+3.1}_{-1.5}$ & $2.0^{+2.4}_{-1.2}$ & $0.1^{+0.2}_{-0.8}$\\
   $\kappa$ & $2.7^{+3.2}_{-3.0}$ & $1.7^{+3.5}_{-3.2}$ & $2.4^{+2.1}_{-2.0}$ & $0.6^{+3.1}_{-2.9}$
\end{tabular}
\end{ruledtabular}
\caption{Merger rate and it's evolution corresponding to each chirp mass interval in Table~\ref{table:peaks_rate_summary}. All units are in $\mathrm{Gpc}^{-3} \mathrm{yr}^{-1}$ and all masses are in $M_\odot$.}
\label{table:peaks_rate_summary}
\end{table}

\begin{figure*}
    \centering
    \includegraphics[width=0.48\textwidth]{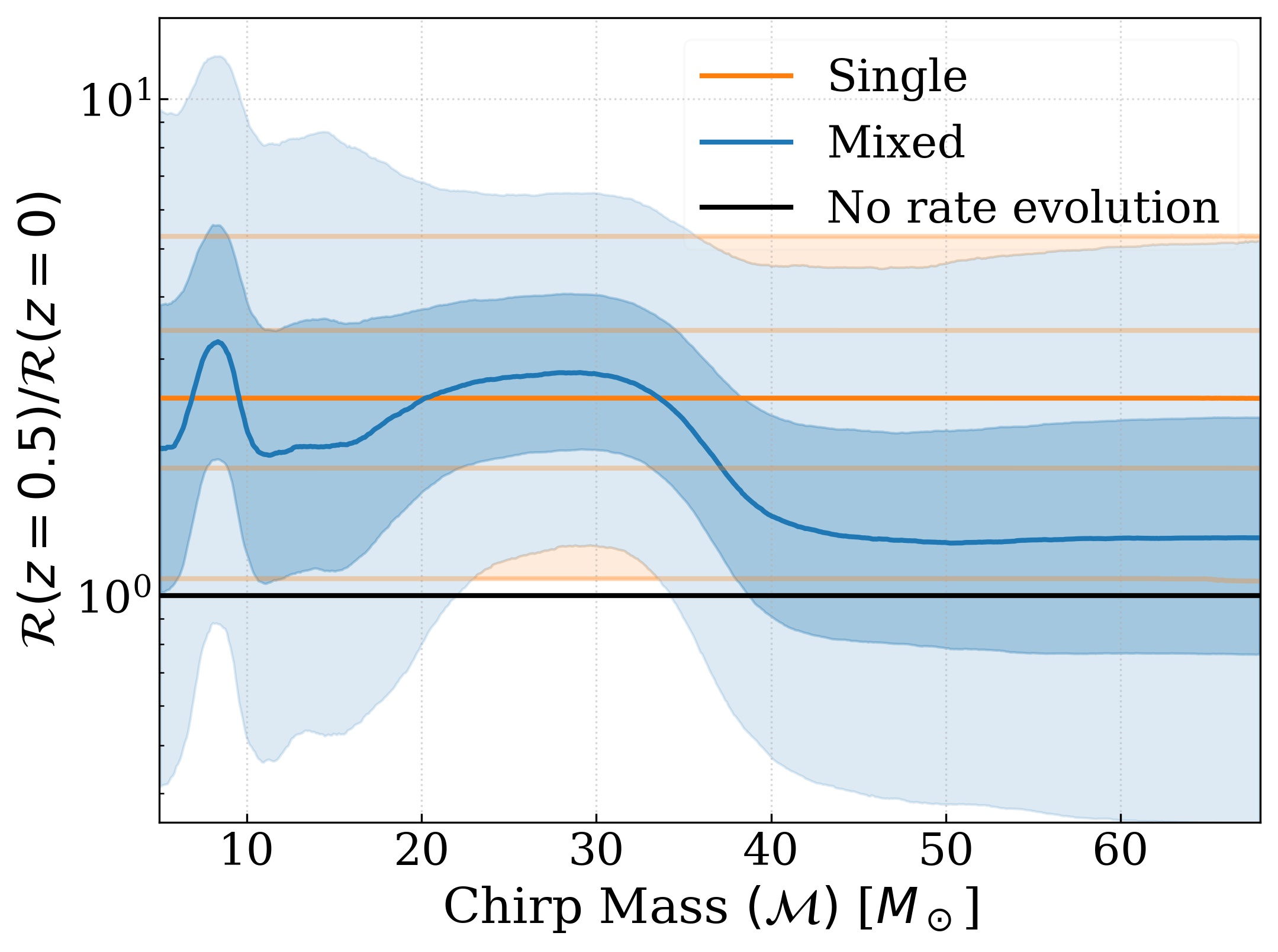}
    \includegraphics[width=0.48\textwidth]{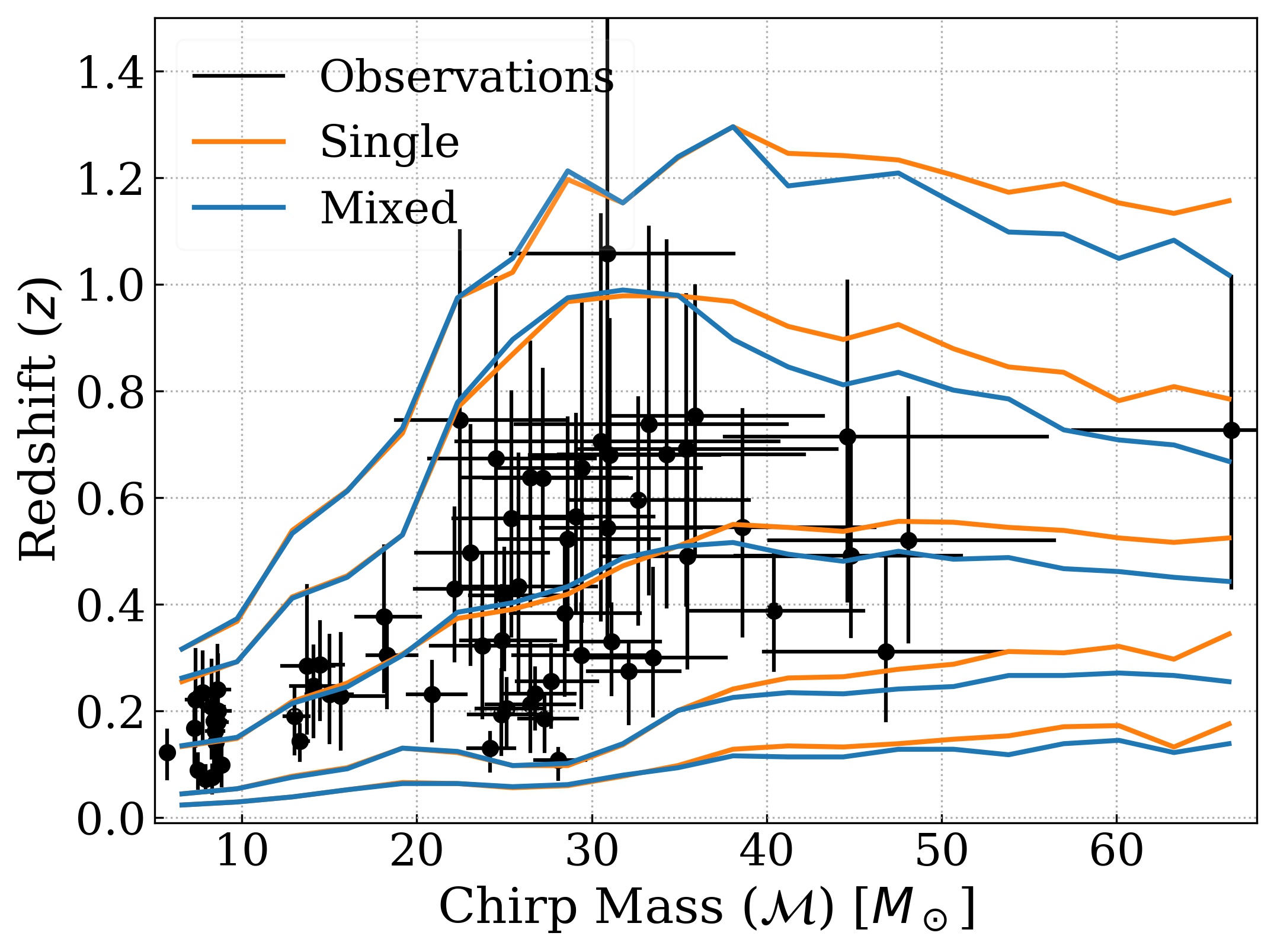}    \caption{\emph{Left)} The dependence of the redshift evolution of the merger rate on the chirp mass. The vertical axes shows the ratio of merger rate between $z = 0$ and $z = 0.5$. The light/dark band (or light solid lines if hidden) is the 90\%/50\% credible interval and the dark solid curve is the median distribution. The blue bands, that correspond to the \emph{Mixed} model, show shallowing of merger rate at the second and the fourth peak. \emph{Right)} Comparing selection weighted prediction with the observations. We apply selection effects to the predicted population and obtain multiple realisations of \emph{expected observations}. We record the minimum, maximum, and median redshift values for each realisation. The top two curves are the $95^{\mathrm{th}}$ and $75^{\mathrm{th}}$ percentile of the maximum values, the bottom two curves are the $5^{\mathrm{th}}$ and $25^{\mathrm{th}}$ percentile of the minimum values and the middle curve is the median of the median values. The black crosses are measurements from the \ac{GW} observations. Both the models predict equivalent distribution for most of the chirp mass range with \emph{Mixture} model making predictions at relatively smaller redshift values for heavier masses.}
    \label{fig:rate_mchirp_zevol}
\end{figure*}

The \emph{Single} model assigns the same redshift evolution to the whole population. For this model, we measure $\kappa = 2.3^{+1.8}_{-2.1}$. The \emph{Mixed} model assigns a separate $\kappa$ for a mixing component and facilitates independent modeling of merger rate evolution for different regions of the population. We recover the mass dependence by marginalising $\kappa$ over the chirp mass intervals listed in Table~\ref{table:peaks_rate_summary}. We report the aggregate $\kappa$'s associated with each peak in Table~\ref{table:peaks_summary}. In addition, similar information is portrayed in Fig.~\ref{fig:rate_mchirp_zevol} where we plot the fractional increase in merger rate from $z=0$ to $z=0.5$ as dependent on the chirp mass. The credible intervals are large but there may be initial hints that the merger rate evolution associated with the peaks are disparate. The lack of observations, especially at higher redshift, for the second and the fourth interval results in shallower evolution of merger rate. In particular, the merger rate evolution for the fourth chirp mass interval is shallower compared to the rest of the \ac{BBH} population at a credibility of 90\%. As mentioned earlier, our inferences are subjected to our choice of the prior on $\kappa$. However, our confidence in this feature increases/decreases on increasing/decreasing the flexibility of the analysis in modeling the rate evolution~(please refer to Section~\ref{sec:analysis} for a discussion on this). The third peak, which is confined in the chirp mass interval 18 to 37$M_\odot$, contributes to more than half of the observed \ac{GW} signals. We do not observe a notable mass evolution in this range as shown in Fig.~\ref{fig:rate_mchirp_zevol}. This figure also shows the predicted observations (selection applied to population prediction), which are consistent for both the models for most of the chirp mass range. However, the \emph{Mixed} model predicts lower redshift values at higher chirp masses.

\subsection{Correlated Features}
\label{sec:correlated}
The mixture model framework allows us to model correlations present among the population's signal parameters (chirp mass, mass ratio aligned spin, or redshift distributions). Once we have obtained the posterior on model hyper-parameters, $\Lambda$, we can predict the population distribution, $p(\theta|\Lambda)$, for the signal parameter, $\theta$. 

\begin{figure*}
    \centering
    \includegraphics[width=0.97\textwidth]{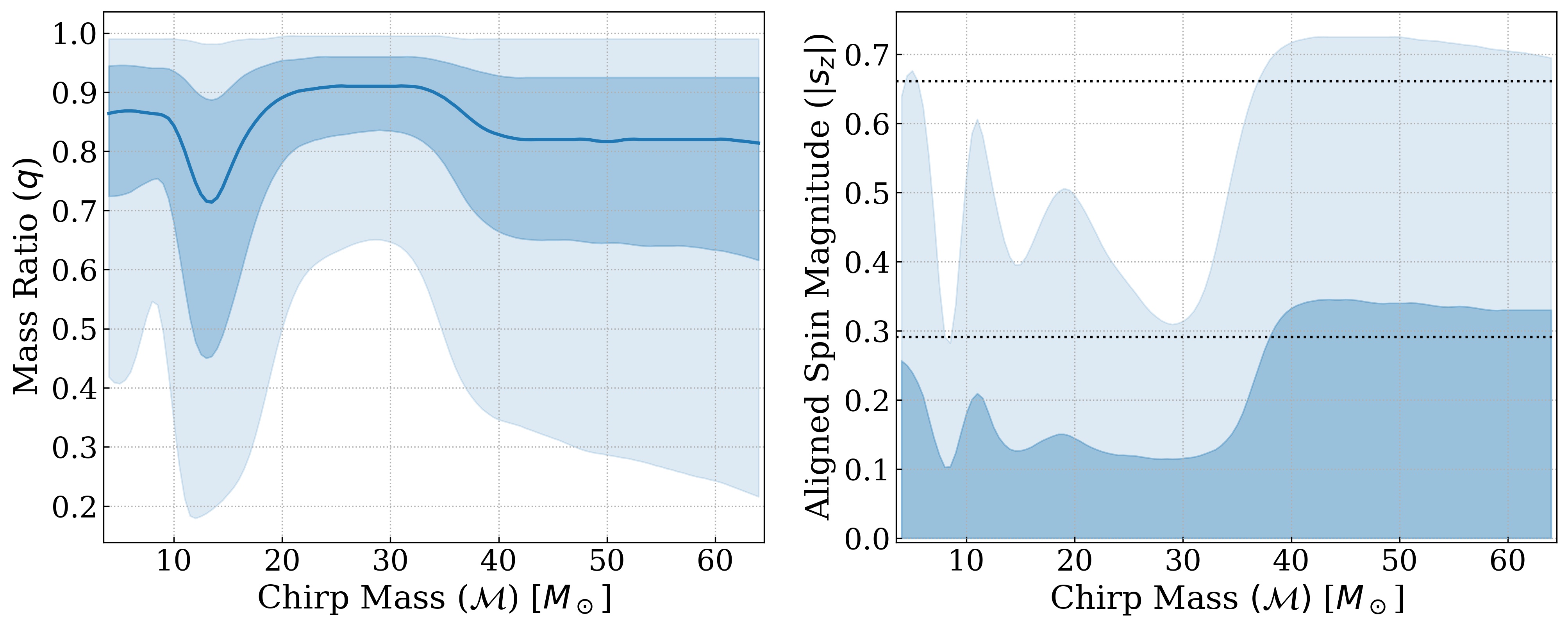}
    \caption{ The variation of mass ratio and aligned spin magnitude with the chirp mass. \emph{Left}) The solid curve is the median and the light/dark bands are the 90\%/50\% credible intervals. Around 95\% of the mergers are consistent with mass ratios of 0.5 or more. The mass ratio show dependence on the chirp mass. Binaries consistent with the second/third peak are most/least asymmetric. \emph{Right}) The light/dark bands are the magnitude of the aligned spin at 90\%/50\% credibility. The aligned spin magnitude is consistent with small values for most of the chirp mass range, however, it increases for chirp mass values of 30 $M_\odot$ or more. The dotted lines are aligned spin's 50\% and 90\% credible bounds of the prior distribution.}
    \label{fig:q_and_sz_vs_mch}
\end{figure*}

\begin{figure}
    \centering
    \includegraphics[width=0.48\textwidth]{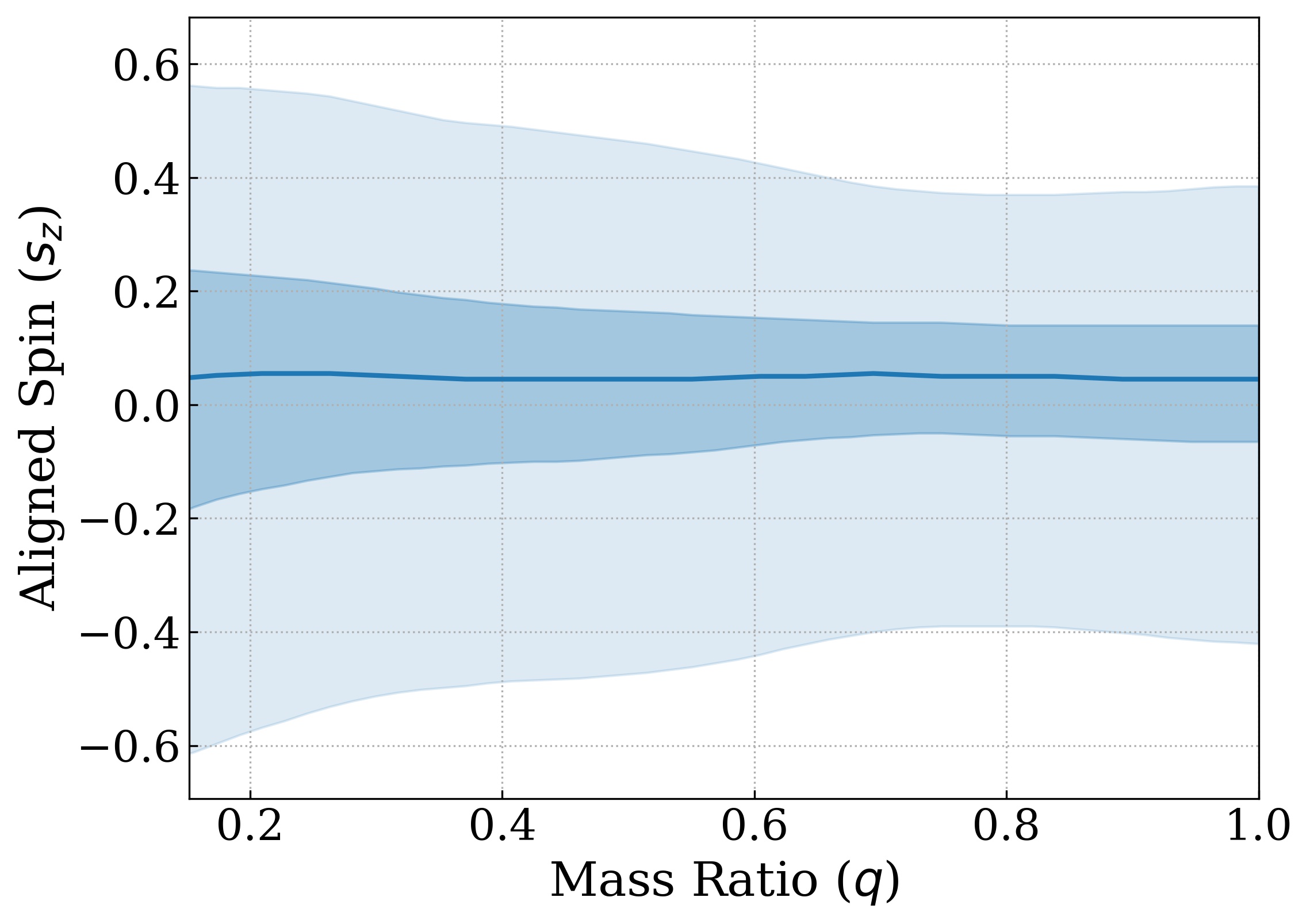}
    \caption{The variation of the aligned spin with the mass ratio. Solid curve is the median distribution, and the light/dark bands are the 90\%/50\% credible intervals. The aligned spin is consistent with very small values and does not seem to show a correlation throughout the mass ratio range.}
    \label{fig:sz_vs_q}
\end{figure}

Fig.~\ref{fig:q_and_sz_vs_mch} shows the variation of the mass ratio and aligned spin as a function of the chirp mass. The mass ratio shows a weak correlation with the chirp mass. Most of the binaries are of comparable masses throughout the chirp mass range. The distribution shows increased asymmetry at the second peak. This is mainly due to GW190412 and multiple observations that have a mass ratio of around one-half. The binaries corresponding to the third peak show least asymmetry. A phenomenological treatment of this feature is reported in \citet{2022arXiv220101905L}. The spins are consistent with small magnitude for most of the chirp mass range but show an increase for chirp masses $30 M_\odot$ or more. The 90\% credible interval for aligned spins averaged over chirp masses 30$M_\odot$ or less is [\lszLomass, \rszLomass], that increases to [\lszHimass, \rszHimass] for chirp masses 30$M_\odot$ or more. Vamana models aligned spin for both the black holes to be independent but identically distributed. We do not observe a correlation between aligned spin and mass-ratio as shown in Fig.~\ref{fig:sz_vs_q} \footnote{Please refer to \citet{2021ApJ...922L...5C} for a phenomenological treatment of correlation between effective spin and mass ratio. This analysis reports these parameters to be anti-correlated.}. The aligned spin distribution is devoid of a trend. We stress that the correlation observed between parameters is seldom independent. A change in an inferred correlation caused due to waveform systematic or change in priors can also lead to a changed inference between other parameters. This is especially relevant for heavier masses where chirp mass, mass ratio, and spin degeneracy are strong.

\section{The Hierarchical Merger Scenario}
\label{sec:hierarchy}

Similar to the suggestion in our previous article~\citep{2021ApJ...913L..19T}, we discuss the observed features in the \ac{BBH} population in the context of hierarchical mergers.

\emph{Peaks and lack of mergers}: In simplest terms, the lack of observations in the chirp mass range 10--12$M_\odot$ and four well-placed peaks can be explained with the first peak populated from binaries that have black holes produced from stellar evolutionary process and following peaks due to hierarchical merger scenario~\citep{2002MNRAS.330..232C, 2016ApJ...831..187A, 2019PhRvD.100d3027R, 2021ApJ...914L..18D, 2021NatAs...5..749G, 2021Symm...13.1678M}. Fig.~\ref{fig:post_mc} suggests, starting at the first peak black holes merge to successively produce heavier black holes. The location of these peaks bears a factor of around 1.9. Such a factor would naturally arise from a hierarchical merger scenario as the remnant produced from the merger of black holes is slightly less massive than the total mass of the binary due to around 5\% loss of mass in gravitational waves.

\emph{Spins}: Arguably, the most robust prediction for a hierarchical merger scenario is highly spinning remnants~\citep{2007ApJ...659L...5C, 2007PhRvL..98i1101G, 2008PhRvD..77b6004B, 2008ApJ...684..822B, 2021PhRvD.104h4002B, 2021ApJ...918L..31M}. Even if we assume that black holes in the first peak/generation to have low spins, the black holes in the second or higher generation should display high spins. And although spins do increase with the masses as seen in Fig.~\ref{fig:q_and_sz_vs_mch} they are consistent with small magnitudes for most of the chirp mass range.

\emph{Cross- generation mergers}: Hierarchical merger scenarios often predict cross generation mergers. Black holes in the first peak/generation(G) can also merge with the heavier ones which are remnant of a previous merger. In fact, many combinations exist resulting in mergers with mass ratios of one-half, one-fourth etc., and giving rise to intermediate peaks located between the primary peaks. Among these, 1G+2G mergers are expected to be most dominant \citep{2019PhRvD.100d3027R, 2020ApJ...894..133A}, but, we are not observing these mergers as their observation will fill the gap in the 10-12$M_\odot$ chirp mass range. However there are some hints of cross-generation mergers. The chirp masses for 1G+3G mergers will overlap with the second peak. GW190412 is an example that has it's chirp and component masses consistent with a 1G+3G merger. In addition, it was observed with moderate spins. Observations GW190408 and GW191215 with chirp mass of around 19 $M_\odot$ are consistent with a 2G+3G merger but are neither observed with a mass-ratio of one-half nor with high spins. However, an emerging peak at this chirp mass value will be of interest.
%
%\section{Sanity Check}
%We perform sanity checks to verify the efficacy of the analyses. Once the posterior on the model hyper-parameters, $\Lambda$, have been obtained it is straightforward to predict the population $p(\theta|\Lambda$. This is also referred as the posterior predictive distribution (PPD). We generated selection weighted PPD by applying selection effects to the predicted population. We also generate various realisations of observed data by randomly selecting from re-weighted posteriors on the estimated signal parameters. The re-weighting essentially adjusts the fiducial prior used in estimating the parameter to the predicted population. Figure \ref{fig:ppc_mc_z} compares the cumulative distributions suggesting the predicted population is robust in predicting the observed data.

%In Figure \ref{fig:obsz} we create a two-dimensional plot for the chirp mass and redshift. We record the minimum, median and maximum values of all the selection weighted PPD's and compare the resulting distributions with the observed data.

\section{Astrophysical Implications}
\label{sec:hiermerge}
The mass spectrum has retained the structure after the addition of new observations. Thus, all of the implications we made earlier still remain valid~\citep{2021ApJ...913L..19T}.

The mass distribution of field binaries is expected to follow a power law like distribution with the maximum mass of the binaries sensitive to the metallicity and the initial mass function of the progenitor stars.  The metallicity of stars impacts the mass loss due to stellar winds~\citep{2003ApJ...591..288H, 2010ApJ...714.1217B, 2014LRR....17....3P}. The mean metallicity decreases with redshift~\citep{2014ARA&A..52..415M}. At lower metallicities, the black hole mass distribution is expected to extend to higher masses. Pair-instability supernovae can impose an upper limit on the maximum mass of the binary as well as introduce a build-up at high masses~\citep{1964ApJS....9..201F, 1967ApJ...150..131R, 1984ApJ...280..825B, 2002ApJ...567..532H}. Thus, population synthesis models that simulate complex physics of stellar evolution expect the maximum black-hole to many tens of solar mass. However, the results presented here provide evidence for a lack of black hole binaries in the chirp mass range 10--12 $M_\odot$. The median differential rate decays by a factor of around 60 in this range. In addition, there is a presence of peaks in the mass distribution which possibly evolve disproportionately with the redshift. We expect these features to be of interest to the population synthesis models.

\ac{BBH} formation and merger can also be facilitated within the star clusters ~\citep{1993Natur.364..423S, 2000ApJ...528L..17P, 2015PhRvL.115e1101R, 2016ApJ...831..187A, 2019PhRvD.100d1301G}. This could include their formation in active galactic nuclei~\citep{2017MNRAS.464..946S, 2020A&A...638A.119G, 2019PhRvL.123r1101Y, 2021ApJ...908..194T, 2021ApJ...920L..42G} or formation of binaries due to scattering in galactic cusps~\citep{2009MNRAS.395.2127O}. The mass spectrum can potentially inform about the many-body dynamics in the star clusters. Specifically, the relative amplitude and the width of the peaks could provide information about the host environment. 

The population has several other features. The spin distribution is consistent with low magnitudes for most of the mass range but heavier binaries also tend to exhibit larger spins. The mass ratio distribution shows a weak dependence on the chirp mass but not on spins. These multifarious population properties are of interest for analysis attempting to explain observation as a mix of multiple formation channels. Many proposed scenarios predict the formation of binaries in a wide mass range~\citep{2010CQGra..27q3001A, 2021arXiv210714239M} and it is possible to estimate contributions from various formation channels that can give rise to the observed distribution~\citep{2021JCAP...03..068H, 2021ApJ...910..152Z, 2021PhRvD.103h3021W, 2021ApJ...913L...5N}. However, more observations are needed to ascertain if a unique combination can give rise to the observed population properties.

The hierarchical merger scenario offers a simple explanation for the location of four well-placed peaks. Our suggestion as hierarchical mergers to be the dominant source for the peaks require addressing a few issues including the issues outlined in Sec.~\ref{sec:hierarchy}. This scenario opens up various other avenues for investigation. The relative location of peaks will quantify the percentage loss of mass in \ac{GW} and will therefore provide an opportunity to test general relativistic predictions of energy emission due to the merger. Although the absolute location of the peaks depends on the assumed cosmology, their relative location should remain unchanged within the framework of standard cosmology. This creates an opportunity to test non-standard cosmological models. Predicted features in the mass spectrum do not provide any non-gravitational information and thus cannot be used to estimate Hubble's constant. But, if a feature can be identified in the source mass-spectrum~\citep{2012PhRvL.108i1101M, 2019ApJ...883L..42F} it is conceivable to conduct a combined test of general relativity and cosmology.

\section{Conclusions}

In this article we reported on the \ac{BBH} population predicted using the observations made during \ac{LVK} first, second, and third observation runs. Endorsing our previous report and corroborating \ac{LVK}'s recent report, we find the mass distribution has four emerging peaks and a lack of mergers in the chirp mass range 10--12$M_\odot$. The population exhibits a small spin magnitude for most of the mass range that increases monotonically for the heavier masses. The mass ratio distribution shows dependence on the chirp mass but not on the aligned spin. We observe possible hints that the redshift evolution of the merger rate is disparate for the peaks in the mass distribution. We expect these features to have large implications on our understanding of the \ac{BBH} formation channels, however, as our results are limited by small statistics we await \ac{LVK}'s fourth observation run which promises to significantly increase the number of observations.

\section*{Acknowledgements}
Sincere thanks to Prof.~Bala Iyer for providing support and guidance during the development of this project.

This work is supported by the STFC grant ST/V005618/1, and European Research Council (ERC) Consolidator Grant 647839.

We are grateful for the computational resources provided by  Cardiff  University and funded by an STFC grant supporting UK Involvement in the Operation of Advanced LIGO. We are also grateful for computational resources provided by the Leonard E Parker Center for Gravitation, Cosmology, and Astrophysics at the University of Wisconsin-Milwaukee and supported by National Science Foundation Grants PHY-1626190 and PHY-1700765

This research has made use of data, software and/or web tools obtained from the Gravitational Wave Open Science Center (https://www.gw-openscience.org/), a service of LIGO Laboratory, the LIGO Scientific Collaboration and the Virgo Collaboration. LIGO Laboratory and Advanced LIGO are funded by the United States National Science Foundation (NSF) as well as the Science and Technology Facilities Council (STFC) of the United Kingdom, the Max-Planck-Society (MPS), and the State of Niedersachsen/Germany for support of the construction of Advanced LIGO and construction and operation of the GEO600 detector. Additional support for Advanced LIGO was provided by the Australian Research Council. Virgo is funded, through the European Gravitational Observatory (EGO), by the French Centre National de Recherche Scientifique (CNRS), the Italian Istituto Nazionale della Fisica Nucleare (INFN) and the Dutch Nikhef, with contributions by institutions from Belgium, Germany, Greece, Hungary, Ireland, Japan, Monaco, Poland, Portugal, Spain.   

\section*{Data Availability}

The code used in performing the presented analysis and the corresponding result files are available at \href{https://github.com/vaibhavtewari/vamana}{https://github.com/vaibhavtewari/vamana}.

\bibliography{references}{}
\bibliographystyle{aasjournal}



\end{document}